\documentclass[aps,preprint,groupedaddress,floatfix]{revtex4}
\usepackage[latin1]{inputenc}
\usepackage[T1]{fontenc}
\usepackage{graphicx}  
\usepackage{dcolumn}   
\usepackage{bm}        
\usepackage{amssymb}   
\usepackage{epsfig}

\begin{document}

\title{IRPF90: a programming environment for high performance computing}

\author{Anthony Scemama}

\affiliation{
Laboratoire de Chimie et Physique Quantiques, CNRS-UMR 5626, \\
IRSAMC Universit\'e Paul Sabatier, 118 route de Narbonne \\
31062 Toulouse Cedex, France
}

\date{\today}

\begin{abstract}
IRPF90 is a Fortran programming environment which helps the development of
large Fortran codes. In Fortran programs, the programmer has to focus on
the order of the instructions: before using a variable, the programmer has to
be sure that it has already been computed in all possible situations. For large
codes, it is common source of error.
In IRPF90 most of the order of instructions is handled by the pre-processor, and
an automatic mechanism guarantees that every entity is built before being used.
This mechanism relies on the \{needs/needed by\} relations between the entities,
which are built automatically. 
Codes written with IRPF90 execute often faster than Fortran programs, are faster
to write and easier to maintain.
\end{abstract}

\pacs{}
\maketitle

\section{Introduction}

The most popular programming languages in high performance computing (HPC) are
those which produce fast executables (Fortran and C for instance). Large
programs written in these languages are difficult to maintain and these
languages are in constant evolution to facilitate the development of large
codes. For example, the C++ language\cite{cpp} was proposed as an improvement
of the C language by introducing classes and other features of object-oriented
programming. In this paper, we propose a Fortran pre-processor with a very
limited number of keywords, which facilitates the development of large programs
and the re-usability of the code without affecting the efficiency.

 In the imperative programming paradigm, a computation is a ordered list of
commands that change the state of the program.  At the lowest level, the
machine code is imperative: the commands are the machine code instructions
and the state of the program is represented by to the content of the memory.
At a higher level, the Fortran language is an imperative language.  Each
statement of a Fortran program modifies the state of the memory.

 In the functional programming paradigm, a computation is the evaluation of a
function. This function, to be evaluated, may need to evaluate other functions.
The state of the program is not known by the programmer, and the memory
management is handled by the compiler.

Imperative languages are easy to understand by machines, while functional
languages are easy to understand by human beings. Hence, code written in an
imperative language can be made extremely efficient, and this is the main
reason why Fortran and C are so popular in the field of High Performance
Computing (HPC).

However, codes written in imperative languages usually become excessively
complicated to maintain and to debug. In a large code, it is often very
difficult for the programmer to have a clear image of the state of the program
at a given position of the code, especially when side-effects in a procedure
modifiy memory locations which are used in other procedures.

In this paper, we present a tool called ``Implicit Reference to Parameters with
Fortran 90'' (IRPF90). It is a Fortran pre-processor which facilitates the
development of large simulation codes, by allowing the programmer to focus on
{\em what} is being computed, instead of {\em how} it is computed. This last
sentence often describes the difference between the functional and the
imperative paradigms\cite{hudak}.
From a practical point of view, IRPF90 is a program written in the
Python\cite{python} language. It produces Fortran source files from IRPF90 source
files. IRPF90 source files are Fortran source files with a limited number of
additional statements.  To explain how to use the IRPF90 tool, we will write a
simple molecular dynamics program as a tutorial.

\section{Tutorial: a molecular dynamics program}

\subsection{Imperative and functional implementation of the potential}
We first choose to implement the Lennard-Jones
potential\cite{lj} to compute the interaction of pairs of atoms:
\begin{equation}
V(r) = 4\epsilon \left[ \left( \frac{\sigma}{r} \right)^{12} -
                        \left( \frac{\sigma}{r} \right)^{6 }  \right]
\end{equation}
where $r$ is the atom-atom distance, $\epsilon$ is the depth of the potential
well and $\sigma$ is the value of $r$ for which the potential energy is zero.
$\epsilon$ and $\sigma$ are the parameters of the force field.

\begin{figure}
\includegraphics{1.irp.f.epsi}
\caption{Imperative implementation of the Lennard-Jones potential.}
\label{fig:irp1}
\end{figure}
Using an imperative style, one would obtain the program given in
figure~\ref{fig:irp1}.  One can clearly see the sequence of statements in
this program: first read the data, then compute the value of the potential.

\begin{figure}
\includegraphics{2.irp.f.epsi}
\caption{Functional implementation of the Lennard-Jones potential.}
\label{fig:irp2}
\end{figure}
This program can be re-written using a functional style, as shown in
figure~\ref{fig:irp2}.  In the functional form of the program, the sequence
of operations does not appear as clearly as in the imperative example.
Moreover, the order of execution of the commands now depends on the choice of
the compiler: the function {\tt sigma\_over\_r} and the function {\tt
epsilon\_lj} are both called on line 12-13, and the order of execution may
differ from one compiler to the other.

\begin{figure}
 \includegraphics{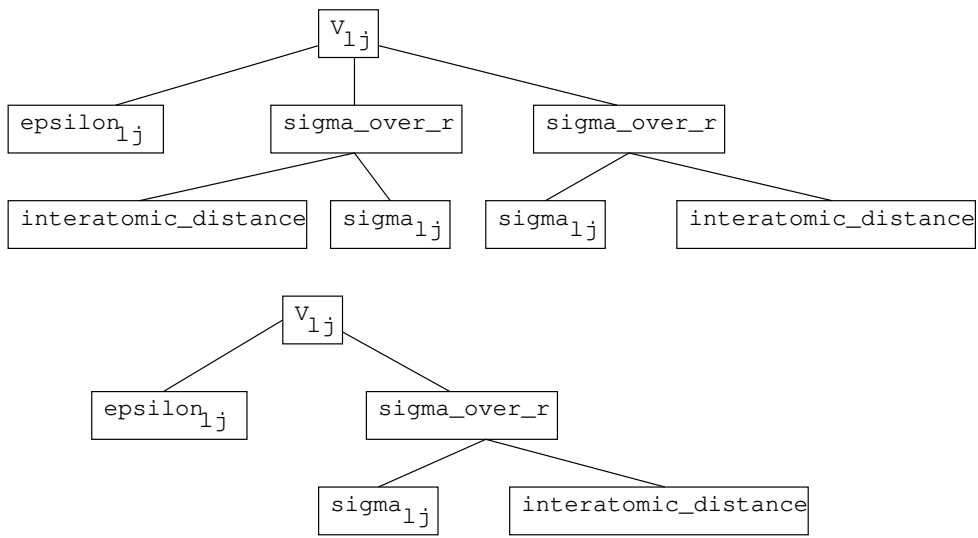}
 \caption{The production tree of {\tt V\_lj}. Above, the tree produced by the
 program of figure~\ref{fig:irp2}. Below, the tree obtained if only one
 call to {\tt sigma\_over\_r} is made.}
 \label{prod_tree}
\end{figure}

The program was written in such a way that the functions have no arguments. The
reason for this choice is that the references to the entities which are needed
to calculate a function appear inside the function, and not outside of the
function. Therefore, the code is simpler to understand for a programmer who
never read this particular code, and it can be easily represented as a
production tree (figure~\ref{prod_tree}, above). This tree exhibits the
relation \{needs/needed by\} between the entities of interest: the entity {\tt
V\_lj} needs the entities {\tt sigma\_over\_r} and {\tt epsilon\_lj} to be
produced, and {\tt sigma\_over\_r} needs {\tt sigma\_lj} and {\tt
interatomic\_distance}.

In the imperative version of the code (figure~\ref{fig:irp1}), the production
tree has to be known by the programmer so he can place the instructions in the
proper order. For simple programs it is not a problem, but for large codes the
production tree can be so large that the programmer is likely to make wrong
assumptions in the dependencies between the entities. This complexifies the
structure of the code by the introduction of many different methods to compute
the same quantity, and the performance of the code can be reduced due to the
computation of entities which are not needed.

In the functional version (figure~\ref{fig:irp2}), the production tree does not
need to be known by the programmer. It exists implicitely through the function
calls, and the evaluation of the main function is realized by exploring the
tree with a depth-first algorithm. A large advantage of the functional style is
that there can only be one way to calculate the value of an entity: calling the
corresponding function.  Therefore, the readability of the code is improved for
a programmer who is not familiar with the program.  Moreover, as soon as an
entity is needed, it is calculated and valid. Writing programs in this way
reduces considerably the risk to use un-initialized variables, or variables
that are supposed to have a given value but which have been modified by a
side-effect.

With the functional example, every time a quantity is needed it is computed,
even if it has already been built before.  If the functions are pure (with no
side-effects), one can implement memoization\cite{memoization,memoization2} to
reduce the computational cost: the last value of the function is saved, and if
the function is called again with the same arguments the last result is
returned instead of computing it again.  In the present example we chose to
write functions with no arguments, so memoization is trivial to implement
(figure~\ref{fig:memo}).
\begin{figure}
\includegraphics{22.irp.f.epsi}
\caption{Memoized {\tt sigma\_over\_r function}}
\label{fig:memo}
\end{figure}
If we consider that the leaves of the production tree
are constant, memoization can be applied to all the functions.
The production tree of {\tt V\_lj} can now be simplified, as shown in
figure~\ref{prod_tree}, below.

\subsection{Presentation of the IRPF90 statements}
IRPF90 is a Fortran pre-processor: it generates Fortran code from source files
which contain keywords specific to the IRPF90 program.
The keywords understood by IRPF90 pre-processor are briefly presented.
They will be examplified in the next subsections for the molecular
dynamics example.

{\tt BEGIN\_PROVIDER ... END\_PROVIDER}\\
Delimitates the definition of a provider (sections~\ref{sub:irp1} and~\ref{sub:irp2}).

{\tt BEGIN\_DOC ... END\_DOC}\\
Delimitates the documentation of the current provider (section~\ref{sub:irp1}).

{\tt BEGIN\_SHELL ... END\_SHELL}\\
Delimitates an embedded script (section~\ref{sub:irp3}).

{\tt ASSERT}\\
Expresses an assertion (section~\ref{sub:irp1}).

{\tt TOUCH}\\
Expresses the modification of the value of an entity by a side-effect
(section~\ref{sub:irp4}).

{\tt FREE}\\
Invalidates an entity and free the associated memory.
(section~\ref{sub:irp6}).

{\tt IRP\_READ / IRP\_WRITE}\\
Reads/Writes the content of the production tree to/from disk (section~\ref{sub:irp5}).

{\tt IRP\_IF ... IRP\_ELSE ... IRP\_ENDIF}\\
Delimitates blocks for conditional compilation (section~\ref{sub:irp5}).

{\tt PROVIDE}\\
Explicit call to the provider of an entity (section~\ref{sub:irp5}).

\subsection{Implementation of the potential using IRPF90}
\label{sub:irp1}

In the IRPF90 environment, the entities of interest are the result of memoized
functions with no arguments. This representation of the data allows its
organization in a production tree, which is built and handled by the IRPF90
pre-processor.  The previous program may be written again using the IRPF90
environment, as shown in figure~\ref{fig:irp3}.
\begin{figure}
\includegraphics{3.irp.f.epsi}
\caption{IRPF90 implementation of the Lennard-Jones potential.}
\label{fig:irp3}
\end{figure}

The program shown in figure~\ref{fig:irp3} is very similar to the functional
program of figure~\ref{fig:irp2}. The difference is that the entities of
interest are not functions anymore, but variables. The variable corresponding
to an entity is provided by calling a providing procedure (or provider),
defined between the keywords {\tt BEGIN\_PROVIDER ... END\_PROVIDER}. In the
IRPF90 environment, a provider can provide several entities (as shown with the
parameters of the potential), although it is preferable to have providers that
provide only one entity.

When an entity has been built, it is tagged as built. Hence, the next call to
the provider will return the last computed value, and will not build the value
again.  This explains why in the IRPF90 environment the parameters of the force
field
are asked only once to the user.

The {\tt ASSERT} keyword was introduced to allow the user to place
assertions\cite{hoare} in the code. An assertion specifies certain general
properties of a value. It is expressed as a logical expression which is
supposed to be always true. If it is not, the program is wrong. 
Assertions in the code provide run-time checks which can dramatically reduce
the time spent finding bugs: if an assertion is not verified, the program 
stops with a message telling the user which assertion caused the program to
fail.

The {\tt BEGIN\_DOC ... END\_DOC} blocks contain the documentation of the
provided entities. The descriptions are encapsulated inside these blocks in
order to facilitate the generation of technical documentation. For each
entity a ``man page'' is created, which contains the \{needs/needed by\}
dependencies of the entity and the description given in the {\tt BEGIN\_DOC
...  END\_DOC} block. This documentation can be accessed by using the 
{\tt irpman} command followed by the name of the entity.

The IRPF90 environment was created to simplify the work of the scientific
programmer. A lot of time is spent creating Makefiles, which describe the
dependencies between the source files for the compilation.
As the IRPF90 tool ``knows'' the production tree, it can build automatically
the Makefiles of programs, without any interaction with the user. When the
user starts a project, he runs the command {\tt irpf90 --init} in an empty
directory. A standard Makefile is created, with the gfortran
compiler\cite{gfortran} as a default. Then, the user starts to write IRPF90
files which contain providers, subroutines, functions and main programs in
files characterized by the {\tt .irp.f} suffix.  Running {\tt make} calls {\tt
irpf90}, and a correct Makefile is automatically produced and used to compile
the code.

\subsection{Providing arrays}
\label{sub:irp2}

Now the basics of IRPF90 are known to the reader, we can show how simple it is
to write a molecular dynamics program. As we will compute the interaction of
several atoms, we will change the previous program such that we produce an
array of potential energies per atom.  We first need to introduce the quantity
{\tt Natoms} which contains the number of atoms.  Figure~\ref{fig:irp4} shows
the code which defines the geometrical parameters of the system.
\begin{figure}
\includegraphics{4.irp.f.epsi}
\caption{Code defining the geometrical parameters of the system}
\label{fig:irp4}
\end{figure} 
Figure~\ref{fig:irp5} shows the providers corresponding to the potential energy $V$
per atom $i$, where it is chosen equal to the Lennard-Jones potential energy:
\begin{equation}
V_i = V^{LJ}_i  = \sum_{j\ne i}^{N_{\rm atoms}} 4 \epsilon \left[
      \left( \frac{\sigma}{||{\bf r}_{ij}||} \right)^{12} -
      \left( \frac{\sigma}{||{\bf r}_{ij}||} \right)^{6 }  \right]
\end{equation}
\begin{figure}
\includegraphics{5.irp.f.epsi}
\caption{Definition of the potential.}
\label{fig:irp5}
\end{figure}
Figure~\ref{fig:irp6} shows the providers corresponding to the kinetic energy $T$
per atom $i$:
\begin{equation}
T_i = \frac{1}{2} m_i ||{\bf v}_i||^2
\end{equation}
where $m_i$ is the mass and ${\bf v}_i$ is the velocity vector of atom $i$. The velocity
vector is chosen to be initialized zero.
\begin{figure}
\includegraphics{6.irp.f.epsi}
\caption{Definition of the kinetic energy.}
\label{fig:irp6}
\end{figure}

The dimensions of arrays are given in the definition of the provider.  If an
entity, defines the dimension of an array, the provider of the dimensioning
entity will be called before allocating the array. This guarantees that the
array will always be allocated with the proper size.  In IRPF90, the memory
allocation of an array entity is not written by the user, but by the
pre-processor.

Memory can be explicitely freed using the keyword {\tt FREE}. For example,
de-allocating the array {\tt velocity} would be done using {\tt FREE velocity}.
If the memory of an entity is freed, the entity is tagged as ``not built'', and
it will be allocated and built again the next time it is needed.

\subsection{Embedding scripts}
\label{sub:irp3}
The IRPF90 environment allows the programmer to write scripts inside his
code. The scripting language that will interpret the script is given in brackets.
The result of the shell script will be inserted in the file, and then will be
interpreted by the Fortran pre-processor. Such scripts can be used to write
templates, or to write in the code some information that has to be retrieved at
compilation. For example, the date when the code was compiled can be inserted
in the source code using the example given in figure~\ref{fig:irp7}.
\begin{figure}
\includegraphics{7.irp.f.epsi}
\caption{Embedded shell script which gets the date of compilation.}
\label{fig:irp7}
\end{figure}

In our molecular dynamics program, the total kinetic energy {\tt E\_kin} is the
sum over all the elements of the kinetic energy vector {\tt T}:
\begin{equation}
E_{\rm kin} = \sum_{i=1}^{N_{\rm atoms}} T_i 
\end{equation}
Similarly, the potential energy {\tt E\_pot} is the sum of all the potential
energies per atom.
\begin{equation}
E_{\rm pot} = \sum_{i=1}^{N_{\rm atoms}} V_i 
\end{equation}
The code to build {\tt E\_kin} and {\tt E\_pot} is very close: only the
names of the variables change, and it is convenient to write the code using a
unique template for both quantities, as shown in figure~\ref{fig:irp8}.
\begin{figure}
\includegraphics{8.irp.f.epsi}
\caption{Providers of the Lennard-Jones potential energy and the kinetic energy
using a template.} \label{fig:irp8}
\end{figure}
In this way, adding a new property which is the sum over all the atomic
properties can done be done in only one line of code: adding the triplet
(Property, Documentation, Atomic Property) to the list of entities at
line 15.

\subsection{Changing the value of an entity by a controlled side-effect}
\label{sub:irp4}

Many computer simulation programs contain iterative processes. In an
iterative process, the same function has to be calculated at each step, but
with different arguments. In our IRPF90 environment, at every iteration the
production tree is the same, but the values of some entities change.  To keep
the program correct, if the value of one entity is changed it has to be tagged
as ``built'' with its new value, and all the entities which depend on this
entity (directly or indirectly) need to be tagged as ``not built''. These last
entities will need to be re-computed during the new iteration.
This mechanism is achieved automatically by the IRPF90 pre-processor using the
keyword {\tt TOUCH}. The side-effect modifying the value of the entity is
controlled, and the program will stay consistent with the change everywhere in
the rest of the code.

In our program, we are now able to compute the kinetic and potential energy of
the system. The next step is now to implement the dynamics. We choose to
use the velocity Verlet algorithm\cite{verlet}:
\begin{eqnarray}
 {\bf r}^{n+1} &=& {\bf r}^n + {\bf v}^n \Delta t + {\bf a}^n \frac{\Delta t^2}{2} \\
 {\bf v}^{n+1} &=& {\bf v}^n + \frac{1}{2}({\bf a}^n + {\bf a}^{n+1})\Delta t
\end{eqnarray}
where ${\bf r}^n$ and ${\bf v}^n$ are respectively the position and velocity
vectors at step $n$, $\Delta t$ is the time step and the acceleration vector
${\bf a}$ is defined as
\begin{equation}
{\bf a} = \sum_{i=1}^{N_{\rm atoms}} -\frac{1}{m_i} \nabla_i E_{\rm pot}
\end{equation}
The velocity Verlet algorithm is written in a subroutine {\tt verlet}, and the
gradient of the potential energy $\nabla E_{\rm pot}$ can be computed by finite
difference (figure~\ref{fig:irp9}).
\begin{figure}
\includegraphics{9.irp.f.epsi}
\caption{Provider of the gradient of the potential.}
\label{fig:irp9}
\end{figure}

Computing a component $i$ of the numerical gradient of $E_{\rm pot}$ can be decomposed in
six steps:
\begin{enumerate}
 \item Change the component $i$ of the coordinate ${\bf r}_i \longrightarrow ({\bf r}_i + \delta)$
 \item Compute the value of $E_{\rm pot}$ 
 \item Change the coordinate $({\bf r}_i + \delta) \longrightarrow ({\bf r}_i - \delta)$
 \item Compute the value of $E_{\rm pot}$ 
 \item Compute the component of the gradient using the two last values of $E_{\rm pot}$
 \item Re-set $({\bf r}_i - \delta) \longrightarrow {\bf r}_i$
\end{enumerate}
The provider of {\tt V\_grad\_numeric} follows these steps: in the internal
loop, the array {\tt coord} is changed (line 16). Touching it (line 17)
invalidates automatically {\tt E\_pot}, since it depends indirectly on {\tt
coord}. 
As the value of {\tt E\_pot} is needed in line 18 and not valid, it is re-computed
between line 17 and line 18.  
The value of {\tt E\_pot} which is affected to {\tt V\_grad\_numeric(k,i)} is the
value of the potential energy, consistent with the current set of atomic coordinates.
Then, the coordinates are changed again (line 19), and the program is informed of
this change at line 20. When the value of {\tt E\_pot} is used again at line 22,
it is consistent with the last change of coordinates.
At line 23 the coordinates are changed again, but no touch statement follows.
The reason for this choice is efficiency, since two cases are possible for the
next instruction: if we are at the last iteration of the loop, we exit the main
loop and line 26 is executed. Otherwise, the next instruction will be line 16.
Touching {\tt coord} is not necessary between line 23 and line 16 since no
other entity is used.

The important point is that the programmer doesn't have to know {\em how} {\tt
E\_pot} depends on {\tt coord}. He only has to apply a simple rule which states
that when the value of an entity $A$ is modified, it has to be touched before
any other entity $B$ is used. If $B$ depends on $A$, it will be re-computed,
otherwise it will not, and the code will always be correct.  Using this
method to compute a numerical gradient allows a programmer who is not
familiar with the code to compute the gradient of any entity $A$ with respect
to any other quantity $B$, without even knowing if $A$ depends on $B$. If $A$
does not depend on $B$, the gradient will automatically be zero.
In the programs dealing with optimization problems, it is a real advantage: a
short script can be written to build automatically all the possible numerical
derivatives, involving all the entities of the program, as given in
figure~\ref{fig:irp11}.
\begin{figure}
\includegraphics{11.irp.f.epsi}
\caption{Automatic generation of all possible gradients of scalar entities with
respect to all other entities.}
\label{fig:irp11}
\end{figure}

The velocity Verlet algorithm can be implemented (figure~\ref{fig:irp10}) as
follows:
\begin{enumerate}
\item Compute the new value of the coordinates
\item Compute the component of the velocities which depends on the old set of
coordinates
\item Touch the coordinates and the velocities
\item Increment the velocities by their component which depends
on the new set of coordinates
\item Touch the velocities
\end{enumerate}
\begin{figure}
\includegraphics{10.irp.f.epsi}
\caption{The velocity Verlet algorithm.}
\label{fig:irp10}
\end{figure}

\subsection{Other Features}
\label{sub:irp5}
\begin{figure}
\includegraphics{101.irp.f.epsi}
\caption{The main program.}
\label{fig:irp101}
\end{figure}
As IRPF90 is designed for HPC, conditional compilation is an essential
requirement. Indeed, it is often used for activating and deactivating blocks of
code defining the behavior of the program under a parallel environment. This is
achieved by the {\tt IRP\_IF...IRP\_ELSE...IRP\_ENDIF} constructs.
In figure~\ref{fig:irp101}, the checkpointing block is activated by running
{\tt irpf90 -DCHECKPOINT}. If the {\tt -D} option is not present, the other
block is activated.

The current state of the production tree can written to disk using the command
{\tt IRP\_WRITE} as in figure~\ref{fig:irp101}. For each entity in the subtrees
of {\tt E\_pot} and {\tt E\_kin}, a file is created with the name of the entity
which contains the value of the entity.  The subtree can be loaded again later
using the {\tt IRP\_READ} statement. This functionality is particularly useful
for adding quickly a checkpointing feature to an existing program.

The {\tt PROVIDE} keyword was added to assign imperatively a {needs/needed by}
relation between two entities. This keyword can be used to associate the value
of an entity to an iteration number in an iterative process, or to help the
preprocessor to produce more efficient code. 

A last convenient feature was added: the declarations of the local variables
do not need anymore to be located before the first executable statement. The
local variables can now be declared anywhere inside the providers, subroutines
and functions.  The IRPF90 pre-processor will put them at the beginning of the
subroutines or functions for the programmer. It allows the user to declare the
variables where the reader needs to know to what they correspond.

\section{Efficiency of the generated code}

In the laboratory, we are currently re-writing a quantum Monte Carlo (QMC)
program, named QMC=Chem, with the IRPF90 tool. The same computation was
realized with the old code (usual Fortran code), and the new code (IRPF90
code). Both codes were compiled with the Intel Fortran compiler version 11.1
using the same options. A benchmark was realized on an Intel Xeon 5140
processor.

The IRPF90 code is faster than the old code by a factor of 1.60: the CPU time
of the IRPF90 executable is 62\% of the CPU time of the old code. This time
reduction is mainly due to the avoidance of computing quantities that are
already computed. The total number of processor instructions is therefore
reduced.

The average number of instructions per processor cycle is 1.47 for the
old code, and 1.81 for the IRPF90 code. This application shows that even
if the un-necessary computations were removed from the old code, the code
produced by IRPF90 would still be more efficient.  The reason is that in
IRPF90, the programmer is guided to write efficient code: the providers are
small subroutines that manipulate a very limited number of memory locations.
This coding style improves the temporal locality of the code\cite{locality1}
and thus minimizes the number of cache misses.

The conclusion of this real-size application is that the overhead due to the
management of the production tree is negligible compared to the efficiency
gained by avoiding to compute many times the same quantity, and by helping
the Fortran compiler to produce optimized code.

\section{Summary}

The IRPF90 environment is proposed for writing programs with reduced
complexity.
This technique for writing programs, called ``Implicit Reference to
Parameters'' (IRP),\cite{francois}
is conform to the recommendations of the ``Open Structure Interfaceable
Programming Environment'' (OSIPE)\cite{osipe}:
\begin{itemize}
 \item Open: Unambiguous identification and access to any entity anywhere in
the program
 \item Interfaceable: Easy addition of any new feature to an existing code
 \item Structured: The additions will have no effect on the program logic
\end{itemize}
The programming paradigm uses some ideas of functional programming
and thus clarifies the correspondance between the mathematical formulas and the
code. Therefore, scientists do not need to be experts in programming to
write clear, reusable and efficient code, as shown with the simple molecular
dynamics code presented in this paper.

The consequences of the locality of the code are multiple:
\begin{itemize}
 \item the code is efficient since the temporal locality is increased,
 \item the overlap of pieces of code written simultaneously by multiple
developers is reduced.
 \item regression testing\cite{regression} can be achieved by writing, for each
entity, a program which tests that the entity is built correctly.
\end{itemize}

Finally, let us mention that the IRPF90 pre-processor generates Fortran 90 which
is fully compatible with standard subroutines and functions. Therefore the produced
Fortran code can be compiled on any architecture, and the usual HPC
libraries (BLAS\cite{blas}, LAPACK\cite{lapack}, MPI\cite{mpi},\dots) can be used.

The IRPF90 program can be downloaded on http://irpf90.sourceforge.net

\section*{Acknowledgments}
The author would like to acknowledge F. Colonna (CNRS, Paris) for teaching him
the IRP method, and long discussions around this subject. The author also would
like to thank P. Reinhardt (Universit\'e Pierre et Marie Curie, Paris) for
testing and enjoying the IRPF90 tool, and F. Spiegelman (Universit\'e Paul
Sabatier, Toulouse) for discussions about the molecular dynamics code.


\end{document}